# OneWeb Satellite Brightness –

# Characterized From 80,000 Visible Light Magnitudes

Anthony Mallama

anthony.mallama@gmail.com

2022 March 10

Abstract

The mean apparent magnitude and the mean of magnitudes adjusted to a standard distance are reported. The illumination phase function for OneWeb satellites is determined and it differs strongly from that of VisorSat spacecraft. Brightness flares are characterized and the mean rate of magnitude variation during a pass is determined. Tools for planning observations that minimize interference from bright satellites are illustrated and discussed.

## 1. Introduction

The many bright satellites in low-earth-orbit are interfering with observational astronomy (Walker and Benvenuti 2022, Hall et al. 2021, Walker et al. 2020a, Walker et al. 2020b, Tyson et al. 2020, Otarola et al. 2020, Gallozzi et al. 2020, Hainaut and Williams 2020, McDowell 2020, Williams et al. 2021a and Williams et al. 2021b). To address this issue the International Astronomical Union (IAU 2022) recently announced the creation of its Centre for the Protection of Dark and Quiet Sky from Satellite Constellation Interference. An important aspect of such protection is characterization of the brightness of these spacecraft. Brightness information can enable researchers to plan their observations with reduced interference from satellites. This paper characterizes the magnitudes of the OneWeb constellation based on an analysis of 80,000 observations acquired by the MMT-9 observatory.

Section 2 describes the hardware, the observations and the database of the MMT-9 robotic observatory. Section 3 reports the mean of OneWeb apparent magnitudes and their dispersion, as well as the corresponding values after adjustment to a standard distance of 1,000 km. Section 4 characterizes the magnitudes by fitting them to a Phase Function (PF) which reduces the uncertainty.

Section 5 describes the application of a physical model called the OneWeb Brightness Function (OBF). The section begins by defining a satellite-centered coordinate system (SCCS) which corresponds to the axes of a OneWeb spacecraft. The SCCS is the frame of reference for the OBF. Then the computation of data needed for the OBF is addressed. The required data includes the positions of the Sun and the MMT-9 in the SCCS system. Next, evaluation of the parameter coefficients for the OBF is described. Finally, the ability of the OBF to improve magnitude predictions by reducing their uncertainties is illustrated.

Section 6 discusses the prediction of satellite magnitudes and illustrates the brightness distribution of OneWeb satellites across the sky. Section 7 characterizes OneWeb flares which are short duration brightness surges and addresses the rate of time varying brightness changes normally seen during a satellite pass. Section 8 lists the limitations of this study. Section 9 discusses the research reported here in the context of other similar studies. The conclusions from this analysis are summarized in Section 10.

Throughout the paper, comparisons are drawn between the brightness characteristics of OneWeb satellites and those of the large constellation of Starlink VisorSat spacecraft being launched by SpaceX. The study of Starlink satellites (Mallama 2021) being used for comparison analyzed more than 100,000 MMT-9 magnitudes and is referred to herein as 'the Starlink paper'.

## 2. MMT-9 hardware and observations

Mini-MegaTORTORA (MMT-9) is located at coordinates 43.650 N and 41.431 E (Beskin et al. 2017). This automated observatory includes nine 71 mm diameter f/1.2 lenses and 2160 x 2560 sCMOS sensors. The detectors record the visible spectrum from blue through red. S. Karpov (private communication) derived a color transformation formula which indicates that MMT-9 magnitudes taken through the clear filter are within 0.1 magnitude of the Johnson-Cousins V band-pass for objects with solar-like color indices. Equations to transform from MMT-9 magnitudes to V-band are provided in the Appendix to this paper.

The [MMT-9 database](#) (Karpov et al. 2015) contains photometry of satellite passes that are collected into *track* files. OneWeb track files containing over 80,000 magnitude records were downloaded for this study. The MMT-9 observations were recorded between February and December 2021 when satellites were more than 20$^o$ above the horizon. The distribution of satellite azimuths relative to that of the Sun was reasonably well distributed as shown in Figure 1.

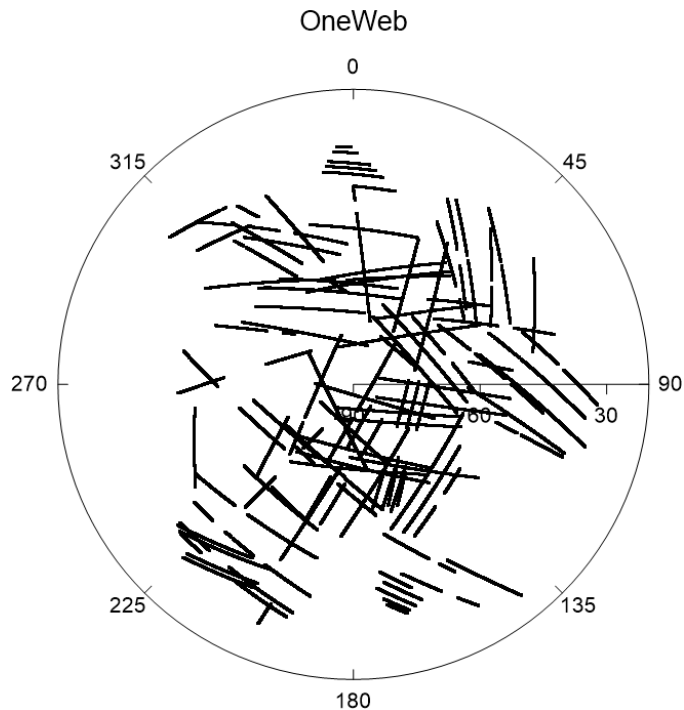

Figure 1. The distribution of satellite tracks in the sky plotted as elevation above the horizon and azimuth relative to the Sun.

The MMT-9 magnitudes include a moderate amount of instrumental scatter. There are 16,052 instances of time-adjacent observations where the same satellite was recorded by different instrument channels within one second of time. The RMS difference of those observation pairs is 0.22 magnitudes.

All spacecraft were at their operational altitude (approximately 1200 km) at the times of the observations according to the [plots](plots) maintained by J. McDowell. Passes containing magnitudes recorded while the satellites were in the Earth's penumbral shadow were excluded.

### 3. Magnitude statistics

Celestial observations are impacted directly by the *apparent magnitudes* of satellites, that is, the values of brightness as recorded by a sensor and corrected only for atmospheric extinction. Table 1 lists mean apparent magnitudes for OneWeb satellites separated into two groups by date. The first group contains data obtained between February and June 2021 and the second group includes July through December observations. The overall mean apparent magnitude is 7.85 with that of the later observations averaging 0.29 magnitudes fainter than the earlier ones. The standard deviations of these values range from 0.68 to 0.71. For the 'scatter removed' standard deviation, the RMS value of 0.22 magnitudes in instrumental dispersion was subtracted in a quadratic sense. The standard deviation of the mean is 0.00 in all cases due to the large number of observations.

Table 1. Apparent magnitudes

| 2021-Months | Mag. | St.Dev. |
|---|---|---|
| 02 – 06 | 7.71 | 0.71 |
| 07 – 12 | 8.00 | 0.68 |
| 02 – 12 | 7.85 | 0.71 |
| Scatter removed | | 0.68 |

In addition to the apparent magnitude, the *magnitude at a standard distance* is used in satellite data analysis. This value is computed by applying the inverse-square law of light to the apparent magnitude.

A distance of 1,000 km is usually chosen as the standard because it is convenient for comparisons between multiple satellite constellations at different altitudes above and below that height.

Table 2 lists 1000-km magnitude statistics for OneWeb satellites with an overall mean of 7.05. The values from the second half of 2021 average 0.22 magnitudes fainter than the earlier data. The standard deviation reduces from 0.71 to 0.58 during that time which indicates that brightness became more predictable during the second half of 2021. This time-dependent aspect of characterization will be revisited in the next section.

Table 2. Magnitudes adjusted to 1,000 km

```
2021-Months   Mag.   St.Dev.

  02 - 06     6.94    0.71
  07 - 12     7.16    0.58

  02 - 12     7.05    0.66
  Scatter removed     0.62
```

The mean apparent brightness of OneWeb satellites is much less than that of VisorSats for which the corresponding average magnitude is 6.43. Meanwhile, at a common distance of 1000 km OneWeb spacecraft are slightly brighter than VisorSats which average 7.21. The VisorSat values are from the Starlink paper.

**4. Phase function analysis**

The previous section reported that standard deviations of OneWeb magnitudes around their mean values are quite large. Hence, these means are of limited use to astronomers who factor satellite brightness into their observing plans. Phase function (PF) analysis offers a potentially more powerful method for quantifying and predicting magnitudes as described below.

Information about satellite illumination based on the geometry of the spacecraft, the Sun and the observer (or sensor) is captured by the *phase angle*. That quantity is the arc length between the Sun and the observer as measured at the spacecraft. For phase angle zero the Sun and observer are aligned as seen from the satellite, and at $180^\circ$ they are in opposite directions.

Satellite brightness computations sometimes use a *standard* PF which corresponds to the fraction of a spherical object illuminated as a function of phase angle. This function is brightest at small phase angles when the side of the satellite facing the observer is mostly lit by the Sun and, conversely, it is faintest when the spacecraft is back-lit.

An improvement over the standard PF is an *empirical* PF which is derived by least-squares fitting of the observed magnitudes at a standard distance to their phase angles. Linear and quadratic equations are commonly used for fitting satellite magnitudes.

The empirical phase function for OneWeb satellites plotted in Figure 2 shows that the spacecraft become fainter with increase phase angle as expected. The linear and quadratic fits are almost identical. The RMS of residuals around the mean linear and quadratic fit lines only reduce to 0.63 in each case, which compares to 0.66 with no fitting as reported in the previous section. After instrumental scatter is removed the RMS is 0.59.

Table 3 lists the linear and quadratic coefficients and RMS values for the entire data set and for the first and second halves of 2021 separately. The RMS values are smaller for the second half of that year, just like the standard deviations of apparent magnitudes and 1000-km magnitudes. Those RMS values are 0.49 magnitudes when instrumental scatter has been removed. The smaller residuals later in the year might indicate that the attitudes (yaw, pitch and roll) of the OneWeb satellites relative to the Sun and observer were controlled in a more consistent manner by that time.

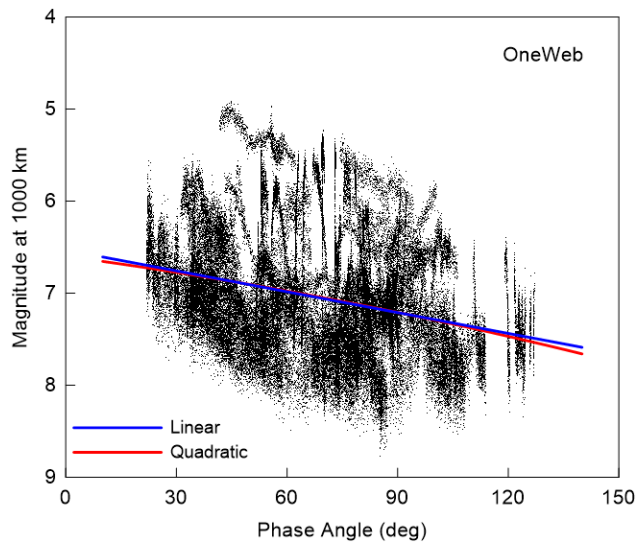

Figure 2. The empirical phase function for OneWeb satellites illustrating linear and quadratic fit functions.

Table 3. Phase function coefficients

```
   2021-Months    -  Linear Degree -      --- RMS ---
                     0        1
      02 - 06       6.505    0.00631       0.69 (0.66)
      07 - 12       6.538    0.00911       0.54 (0.49)
      02 - 12       6.531    0.00755       0.63 (0.59)

   2021-Months    ---- Quadratic Degree ----   --- RMS ---
                     0        1         2
      02 - 06       6.606    0.00323   0.0000209   0.69 (0.66)
      07 - 12       6.799   -0.00016   0.0000715   0.54 (0.49)
      02 - 02       6.600    0.00531   0.0000161   0.63 (0.59)

   Coefficients apply to angles in degrees
   RMS values in parentheses have instrumental scatter removed
```

While empirical PF analysis does not greatly improve the accuracy of magnitude prediction for OneWeb satellites it is useful to know that their behavior is much different from the standard PF and from the PF for VisorSats. This comparison is shown in Figure 3. The highly curved PF for VisorSat indicates strong forward and backward reflection of sunlight. This behavior is not seen with OneWeb satellites. In fact, the directional response of OneWeb is even weaker than that indicated by the standard PF.

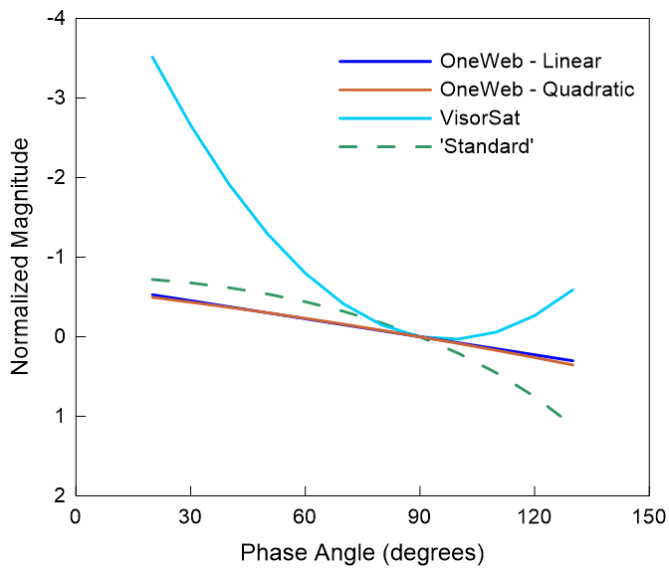

Figure 3. The linear and quadratic PFs for OneWeb along with the VisorSat and 'standard' PFs. All functions are normalized to magnitude 0 at phase angle 90$^o$.

## 5. OneWeb Brightness Function

A physical model for satellite illumination was described in the Starlink paper and here it is applied to OneWeb spacecraft. In this section, a satellite-centered coordinate system (SCCS) corresponding to the general shape and orientation of a satellite is defined. The SCCS serves as the frame of reference for the OneWeb Brightness Function (OBF). Then the computation of data needed for the OBF is detailed. This includes the positions of the MMT-9 observatory and the Sun in the SCCS. Subsequently, evaluation of the parameter coefficients for the OBF is described. Finally, the ability of the OBF to improve magnitude predictions by reducing their uncertainties is demonstrated.

According to the model shown here, https://en.wikipedia.org/wiki/OneWeb#/media/File:KSC-20170316-PH_KLS01_0007~orig.jpg, OneWeb satellites have a boxy shape with a pair of large solar panels on protruding arms. This geometrical shape is the motivation for the SCCS. The north pole of the SCCS is in the direction of the satellite zenith and the zero point of azimuth is toward the Sun as illustrated in Figure 4.

The critical directions for any physical brightness model are those of the lighting source and of the sensor. For the OneWeb magnitudes analyzed here those directions relate to the Sun and the MMT-9 observatory. The components of the direction vectors are azimuth and elevation, and they are distinct from those of the ground-based coordinate system. The elevations of the Sun and of MMT-9 are measured relative to the zenith of the SCCS. The azimuth of the SCCS is defined to be zero in the solar direction. So, the azimuth of MMT-9 is taken to be the difference from that of the solar azimuth.

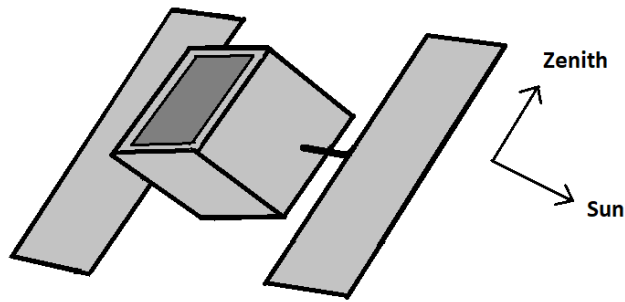

Figure 4. The SCCS frame of reference is illustrated in this schematic of a OneWeb spacecraft. The north pole direction (elevation 90$^o$) is toward zenith. Solar and MMT-9 elevations are generally negative as measured from zenith for satellites visible from the ground. The plane of the solar panels is perpendicular to the Sun, and the solar direction marks the zero point of azimuth.

Coordinates of the Sun and the MMT-9 site in the SCCS were computed as follows. Two line element sets (TLEs) for the spacecraft orbit at the time of observation were acquired from Space-Track. Next, the spacecraft Cartesian coordinates in the Earth-Centered Earth-Fixed reference frame were generated using SPG4 software. The corresponding right ascension (RA) and declination (Dec) values for the north pole of the SCCS were calculated and then adjusted for precession to the time of the observation. The Cartesian positions of the MMT-9 site were computed from its longitude, latitude and height, together with the sidereal time. The precession-adjusted Cartesian coordinates of the satellite were subtracted from the MMT-9 coordinates to determine the RA and Dec of the site which, in turn, were transformed to its azimuth and elevation in the SCCS. The solar RA and Dec and the sidereal time were acquired from the JPL Horizons ephemeris system. Solar coordinates were mapped to azimuth and elevation in the SCCS. Lastly, the MMT-9 azimuth difference was calculated by subtracting the solar azimuth.

Coefficients of the three OBF parameters (solar elevation, MMT-9 elevation and MMT-9 azimuth) were determined by fitting them to the 1,000-km magnitudes. The fitting process was iterated 100 times and the resulting coefficients are listed in Table 4. Subsections of the table correspond to 'all data' and to the first and second halves of 2021 taken separately. Values of the RMS between the observations and the OBF are also given in the Table. The RMS values during the second half of 2021 are smaller than those in the first half. When instrumental scatter is removed from the later data the RMS reduces to 0.44 which is less than that of the empirical PF.

Table 4. Coefficients of the OBF parameters

```
Months 02-06          ------ Polynomial Degree ------   --- RMS ---
Parameter                0          1           2        0.65 (0.62)
---------              ------   ---------   ----------
Azimuth                 7.121    -0.00808    0.0000484
Solar elevation         0.728     0.03232
MMT elevation          -0.920    -0.01416

Months 07-12          ------ Polynomial Degree ------   --- RMS ---
Parameter                0          1           2        0.49 (0.44)
---------              ------   ---------   ----------
Azimuth                 6.579     0.01372   -0.0000592
Solar elevation         0.514     0.02320
MMT elevation           0.648     0.01047
```

```
All data                    ------ Polynomial Degree ------    --- RMS ---
Parameter                    0           1            2         0.61 (0.57)
---------                  ------    ---------    ----------
Azimuth                     6.876      0.00299    -0.0000085
Solar elevation             0.768      0.03439
MMT elevation              -0.113     -0.00177

Coefficients apply to angles in degrees
RMS values in parentheses have instrumental scatter removed
```

The solar elevation function for OneWeb is plotted in Figure 5 which shows that brightness increases as elevation becomes more negative. This geometry corresponds to sunlight impacting the nadir-facing side of the spacecraft at a more perpendicular angle. Brightening can be expected in this case because the nadir side also faces the observer.

The MMT-9 azimuth function is illustrated in Figure 6. This relationship resembles that of the PF (see Figures 2 and 3) because the geometries are somewhat similar. In both cases a small value of the independent variable correlates with a large angular separation between the satellite and the Sun as seen from the MMT-9 site, and vice-versa.

The graph of OneWeb brightness versus MMT-9 elevation indicates very little correlation as shown in Figure 7. The boxy shape of the satellites does not lend itself to increased brightness at any particular viewing angle.

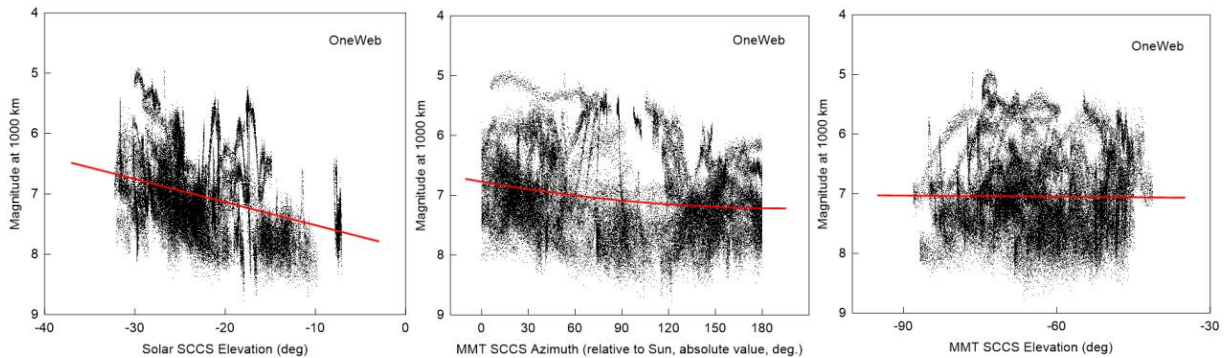

Figures 5, 6 and 7. OneWeb satellite brightness plotted as a function of solar elevation, MMT-9 azimuth and MMT-9 elevation in the SCCS system, respectively.

## 6. Magnitude prediction

Astronomers can avoid the brightest satellites by selecting favorable sky areas and optimal times of night for acquiring observations. OneWeb PF and OBF models used to *characterize* satellite brightness can also *predict* 1000-km magnitudes. Those 1000-km magnitudes can, in turn, can be converted to apparent magnitudes by applying the inverse-square law of light. The inputs needed to compute model brightness are the azimuths and elevations of the Sun and of the spacecraft in the local horizon reference frame as well as the satellite distance.

Examples of brightness predictions for OneWeb satellites based on the PF and OBF are shown as all-sky maps in Figures 8 and 9, respectively. Apparent magnitudes at the end of astronomical twilight with the Sun $18^o$ below the horizon are represented by colors. Both models indicate that the brightest satellites (magnitude < 8.0, shown in red) are located near zenith and slightly offset toward the azimuth that is opposite to the Sun. Fainter satellites are represented by other color bands and the Earth shadow region is shown in black. The brightness distribution for VisorSat spacecraft (from the Starlink paper) is shown in Figure 10 for comparison. Notice that the distribution for VisorSat is shaped differently and that the magnitudes are brighter.

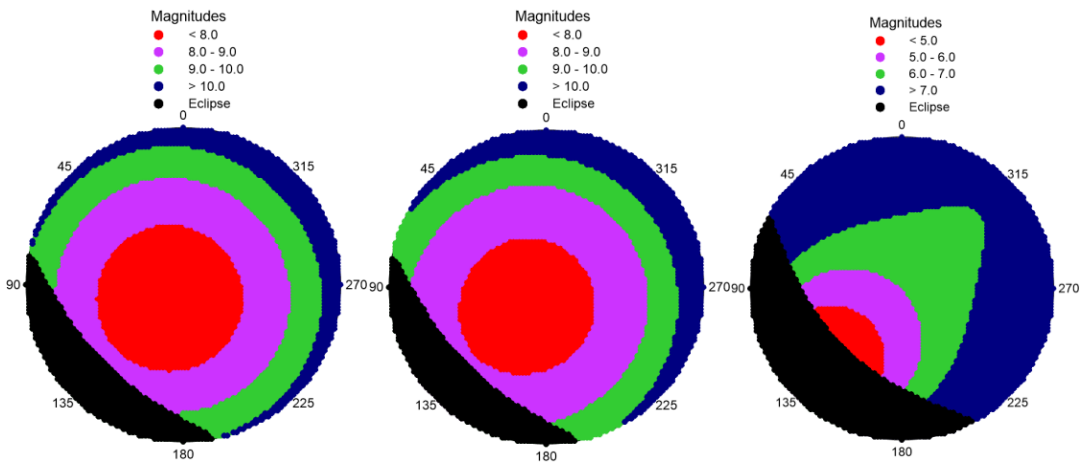

Figures 8, 9 and 10. OneWeb apparent magnitudes over the whole sky as predicted by the PF (left) and the OBF (center) models. VisorSat PF magnitudes are shown at right. Notice that the color legend for OneWeb is fainter by three magnitudes. The Sun is located $18^o$ below azimuth $315^o$.

Model magnitude predictions are also useful for generating brightness statistics and understanding changes that occur with time of night. Figure 11 plots the percentage of sky occupied by OneWeb satellites brighter than three threshold magnitudes. There are separate sets of curves for the whole sky and for that part above 30° elevation. The graph indicates that interference from OneWeb spacecraft brighter than magnitude 9 and above 30° elevation peaks when the Sun is between 20° and 30° below the horizon. If the Sun is nearer the horizon than 20° the satellites tend to be fainter and if the Sun is further from the horizon than 30° many of the spacecraft are eclipsed by the Earth's shadow.

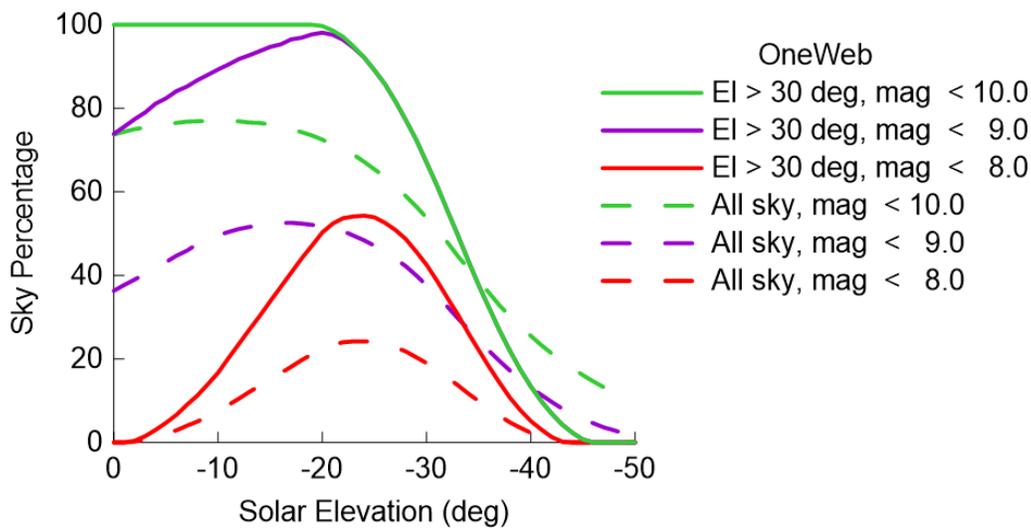

Figure 11. The percentage of sky area with OneWeb magnitudes brighter than magnitudes 8, 9 and 10, is plotted as a function of solar elevation. The solid lines are for the sky above 30° elevation while the dashed lines correspond to the whole sky. The data were generated with the OBF model.

The high 1,200 km altitude of OneWeb satellites makes them fainter the VisorSats at 550 km, which is more favorable for astronomical observers. However, the lower VisorSat altitudes have the advantage that they are eclipsed more than OneWeb spacecraft. Figures 12 and 13 illustrate the different sizes of the eclipse regions when the Sun is at -30° elevation.

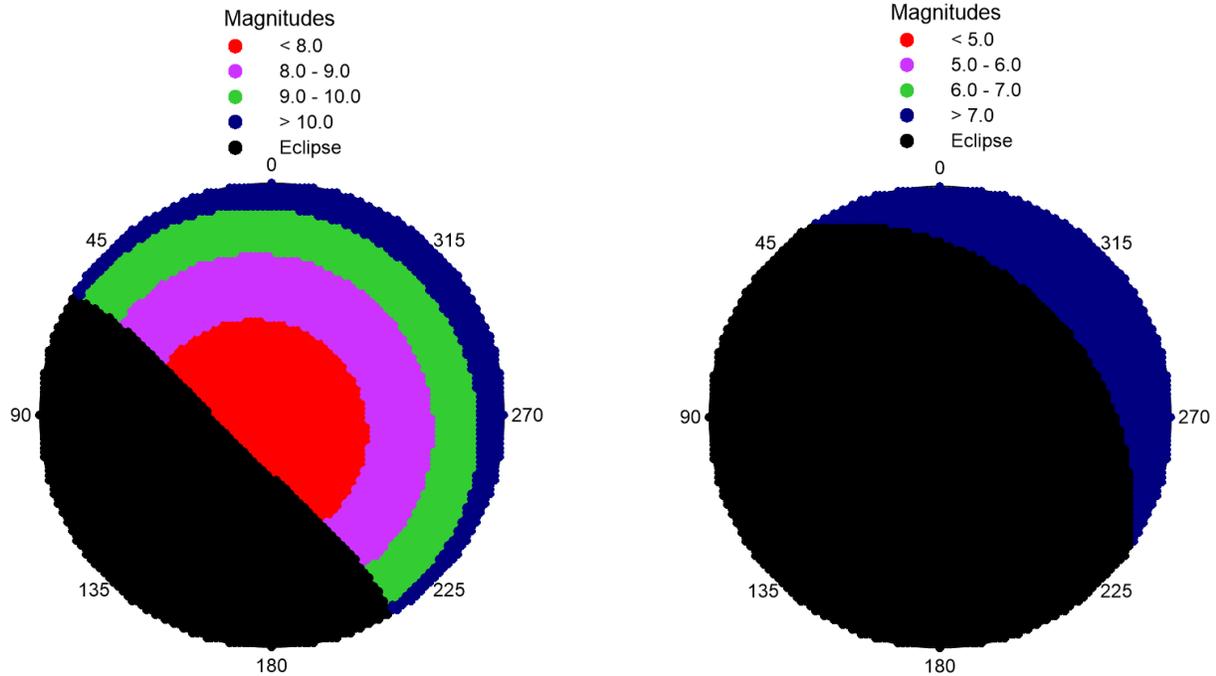

Figures 12 and 13. The eclipse region for VisorSats (right) includes nearly the whole sky when the Sun is 30° below the horizon. However, that region covers less than half of the sky for OneWeb satellites (left) at the same solar elevation.

## 7. Brightness changes on short time scales

The intensity of trails from OneWeb satellites on individual astronomical images will change with time and with position on the sensor. Statistics of these variations may be useful in developing algorithms to reduce the impact of trails and to minimize the loss of science data. So, this section describes magnitude changes that occur on time-scales from seconds to tens of seconds.

The light curve graphs used for the analysis in this paper were inspected by eye for sudden brightness enhancements, called flares. Figure 14 shows an example of one of these surges. Flares are probably due to specular reflection of sunlight from polished surfaces on the satellites.

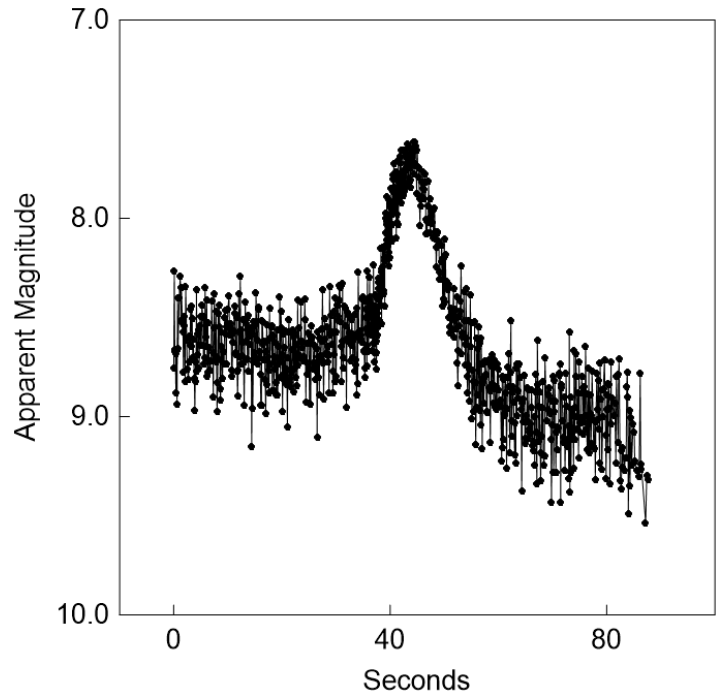

Figure 14. Flaring of OneWeb-0080 recorded on 2021 August 30.

The mean interval between flares that exceeded amplitude 0.5 in apparent magnitude was 721 seconds. This is much less frequent than VisorSat flares for which the mean interval is 129 seconds according to the Starlink paper. The percentage of the elapsed time in all the light curves spent in flares above 0.5 magnitudes was 0.9% for OneWeb as compared to 3.9% for VisorSat.

More gradual brightness variations during a pass were characterized by taking the means of apparent magnitudes in 10 second time intervals. The averaging was performed to reduce instrumental scatter. Mean magnitudes in the adjacent time intervals were then differenced, and that result was divided by the time interval. The absolute value of the quotient was taken to be the rate of change per second. Light curves containing flares were not used in this analysis.

The 10-second averages were also used to determine rates of magnitude change per degree of arc distance that the satellite traveled. This rate was determined by factoring in the satellite's azimuth and elevation at the mid-point of each interval.

The mean rates of OneWeb magnitude variation are given in Table 5 along with standard deviations of the dispersions and standard deviations of the means. The mean rates of change are 0.011 magnitudes per second and 0.040 magnitudes per degree. The standard deviations are nearly as large as the means, which indicates that the rates are highly variable. Nevertheless, the standard deviations of the means are very small, so the mean values are reliable. The Table also lists the corresponding statistics for VisorSat, taken from the Starlink paper. The VisorSat rate of change per second is nearly twice as large as that for OneWeb but the rate of change per degree is about that same size.

Table 5. Gradual brightness variations

|  | Per Second | Per Degree |
|---|---|---|
| **OneWeb** | | |
| Mean rate | 0.011 | 0.040 |
| S.D. | 0.010 | 0.038 |
| S.D. of mean | 0.001 | 0.003 |
| **VisorSat** | | |
| Mean rate | 0.021 | 0.038 |
| S.D. | 0.017 | 0.028 |
| S.D. of mean | 0.002 | 0.004 |

## 8. Limitations of this study

Several factors limit the general applicability and the accuracy of the results from this study. To begin with, the physical modeling of the spacecraft shape is rudimentary and there is no accounting for the light scattering properties of the surfaces.

Additionally, the satellite bodies and solar arrays are only *nominally* aligned with the SCCS frame of reference, while their *actual* attitude (yaw, pitch and roll) at the time of each observation is unknown. Thus the SCCS angles may not be exactly aligned with the OBF parameters corresponding to the physical spacecraft. As a result, the SCCS and OBF do not constitute a formal bi-directional reflectance model.

A further limitation is that the data are all from one location. If spacecraft attitude depends upon geographic latitude and longitude then the magnitudes acquired at the MMT-9 site may not be representative of other regions.

Finally, the observations only span 11 months and there is some evidence for brightness change during that interval. Thus, the observed magnitude of OneWeb satellites could evolve in the future.

## 9. Discussion

This section puts the analysis reported here into the context of other photometric research involving OneWeb satellites. The first such study (Mallama, 2020) was a small-scale investigation of 639 MMT-9 magnitudes that were acquired shortly after the first satellites reached their operational altitude. The mean 1000-km magnitude from that examination was 7.18 as compared to 7.05 reported here. That paper did not report statistics pertaining to the satellite phase function, flaring frequency or rates of brightness change.

Scott et al. (2021) reported on a space-based photometric study of satellite constellations. The mean of their 'open filter' OneWeb magnitudes at the 1,000 km distance was 7.4, or about 0.4 magnitude fainter than that reported here and 0.2 fainter than Mallama (2020).

Krantz et al. (2021) report considerably fainter results. They give V-band magnitudes 'as-observed, only corrected for airmass extinction' which are like the apparent magnitudes listed in this paper. Their mean of 9.1 compares to 7.85 as reported here, so there is a great difference between the results of the two studies.

Krantz et al. explain their faint values as being due to 'an increased number of measurements low in elevation' and point out that 'more satellites are visible low in elevation'. These two points can account for the dim magnitudes resulting from their all-sky observing strategy. While their paper does not report an average satellite elevation, the geometry of a hemisphere gives some indication. Half the surface area of that geometric shape is above $30^o$ and the other half is below. This suggests a value for the mean elevation of an all-sky survey. By contrast, the mean satellite elevation for the photometry reported in this paper is considerably higher at $57^o$.

## 10. Conclusions

More than 80,000 OneWeb magnitudes recorded by the MMT-9 robotic observatory were analyzed. These visible light values represent brightness from the red part of the spectrum through the blue, and they are within about 0.1 magnitude of the V-band.

The mean apparent magnitude for OneWeb satellites is 7.85 with a standard deviation of 0.68 after the estimated instrumental scatter is removed. The mean of magnitudes adjusted to a range of 1,000 km is 7.05 with a standard deviation of 0.62 when adjusted for the scatter. The formal uncertainties of these means are less than 0.01 magnitudes. There is some evidence that the satellites became fainter during the second half of year 2021.

The phase function (PF) for OneWeb satellites indicates that brightness is only weakly related to this illumination angle. The RMS residuals for a linear fit and a quadratic fit are the same. The RMS to the PF for all of 2021 with instrumental scatter removed is 0.59 and for the latter half of the year alone it is 0.49.

A OneWeb Brightness Function (OBF) was tailored to the shape and orientation of the satellites. The OBF fits the observed magnitudes with an RMS residual of 0.57 for the whole year of 2021 after instrumental scatter is removed and it reduces to 0.44 for the latter half of the year. So, the OBF represents OneWeb magnitudes more accurately than does the PF.

OneWeb flares that exceed amplitude 0.5 in apparent magnitude occur once per 721 seconds on average. Such flaring occupies 0.9% of the elapsed time. The average rates of gradual brightness variation are 0.011 magnitudes per second and 0.040 magnitudes per degrees. These gradual rates do not include flares.

The mean of 1000-km OneWeb magnitudes found in this study are reasonably consistent with those derived by Mallama (2020) and by Scott et al. (2021). However, Krantz et al. (2021) report a much fainter mean apparent magnitude. That difference is probably due to the lower elevations of the satellites sampled by Krantz et al.

Two aspects of planning astronomical observations to reduce interference from OneWeb satellites are discussed. One is the use of sky maps showing satellite brightness as a function of azimuth and elevation. The other is a graph showing the percentage of satellite magnitudes above threshold values as a function of the solar elevation.

**Appendix. MMT-9 and V-band magnitude transformation**

A private communication from S. Karpov indicated that '[MMT-9] Instrumental ("Clear") magnitudes are in the system which is roughly Johnson V + 0.15 * (B-V)'. So,

$$V = M_C - 0.15 * (B-V)$$

Equation A-1

where $M_C$ is the MMT-9 clear magnitude and *B-V* is the blue-minus-visual color index. For the solar color index of 0.63 this evaluates to,

$$V = M_C - 0.09.$$

Equation A-2

Thus, the MMT-9 clear magnitude is within about 0.1 magnitude of the V-band for grey bodies reflecting sunlight.